\documentclass[10pt,prb,aps,superscriptaddress,twocolumn,longbibliography]{revtex4-1}
\usepackage{amssymb,amsmath}
\usepackage{graphicx}
\usepackage{color}
\usepackage{natbib}
\usepackage{hyperref}
\usepackage[utf8]{inputenc}

\graphicspath{{figures/}}

\newcommand{\eps}{\varepsilon}      %Greek epsilon
\newcommand{\om}{\omega}      %Greek omega
\usepackage{amsmath}
\usepackage{graphicx}
\usepackage[utf8]{inputenc}
\usepackage{color}
\usepackage{braket}

\begin{document}
\renewcommand{\cite}[1]{[\onlinecite{#1}]}
\title{Photonic topological states mediated by staggered bianisotropy}

\author{Daniel A. Bobylev}
\affiliation{ITMO University, Saint Petersburg 197101, Russia} 

\author{Daria A. Smirnova}
\affiliation{Nonlinear Physics Centre, Australian National University, Canberra ACT 2601, Australia}
\affiliation{Institute of Applied Physics, Russian Academy of Science, Nizhny Novgorod 603950, Russia}

\author{Maxim~A.~Gorlach}
\email{m.gorlach@metalab.ifmo.ru}
\affiliation{ITMO University, Saint Petersburg 197101, Russia} 
 
\begin{abstract}
Photonic topological structures supporting spin-momentum locked topological states underpin a plethora of prospects and applications for disorder-robust routing of light. 
One of the cornerstone ideas to realize such states is to exploit uniform bianisotropic response in periodic structures with appropriate lattice symmetries, which together enable the topological bandgaps. 
Here, it is demonstrated that staggered bianisotropic response gives rise to the topological states even in a simple lattice geometry whose counterpart with uniform bianisotropy is topologically trivial. 
The reason behind this intriguing behavior is in the difference of the effective coupling between the resonant elements with the same and with the opposite signs of bianisotropy. 
Based on this insight, a one-dimensional equidistant array is designed, which consists of high-index all-dielectric particles with alternating signs of bianisotropic response. 
The array possesses chiral symmetry and hosts topologically protected edge states pinned to the frequencies of hybrid magneto-electric modes. These results pave a way towards flexible engineering of topologically robust light localization and propagation by encoding spatially varying bianisotropy patterns in photonic structures.
\end{abstract}

\maketitle

\section{Introduction}\label{sec:Intro}

Topological photonics harnesses nontrivial topology of photonic bands to engineer unidirectional edge states protected against backscattering on defects and robust to fabrication imperfections~\cite{Lu2014,Lu2016,Khanikaev17,Ozawa_RMP}. First proposals of photonic topological structures~\cite{Raghu-PRA-2008,Haldane-PRL-2008,Wang-Soljacic} relied on breaking of time-reversal symmetry which is challenging to achieve in the infrared and optical domains. The alternative strategy, first proposed in Ref.~\cite{Khanikaev}, is to exploit time-reversal-invariant metamaterials with topological helical edge states enabled by bianisotropic response.
 
Bianisotropy, or magneto-electric coupling, couples incident magnetic field to the electric dipole moment of the particle and electric field to its magnetic dipole moment. It can be viewed as a photonic analogue of spin-orbit coupling in condensed matter systems~\cite{Whittaker}. Bianisotropy can be introduced into single particle with overlapped magnetic and electric dipolar resonances by breaking its spatial inversion symmetry~\cite{Serdyukov}. 
Recent works suggest that the concept of bianisotropic topological metamaterials can be applied throughout the entire electromagnetic spectrum, starting from microwaves~\cite{Slob-Scirep,Slob-NP,Slob-2019,Yang-Chen} with a potential of further downscaling towards infrared and optical domains~\cite{Kruk2018,Sievenpiper}.

However, the previous studies~\cite{Chen-Chan,Slob-NP,Ma-Khanikaev,Yang-Chen,Sievenpiper,Cheng-Khanikaev,Slob-Scirep,Slob-2019,AGorlach} focused only on the case of uniform bianisotropy which was used as a mechanism to open  topologically nontrivial bandgap. Specifically, those works investigated either edge or interface states at the boundary between the two domains with the opposite signs of magneto-electric coupling.

\begin{figure}[b]
\begin{center}
\includegraphics[width=1.0\linewidth]{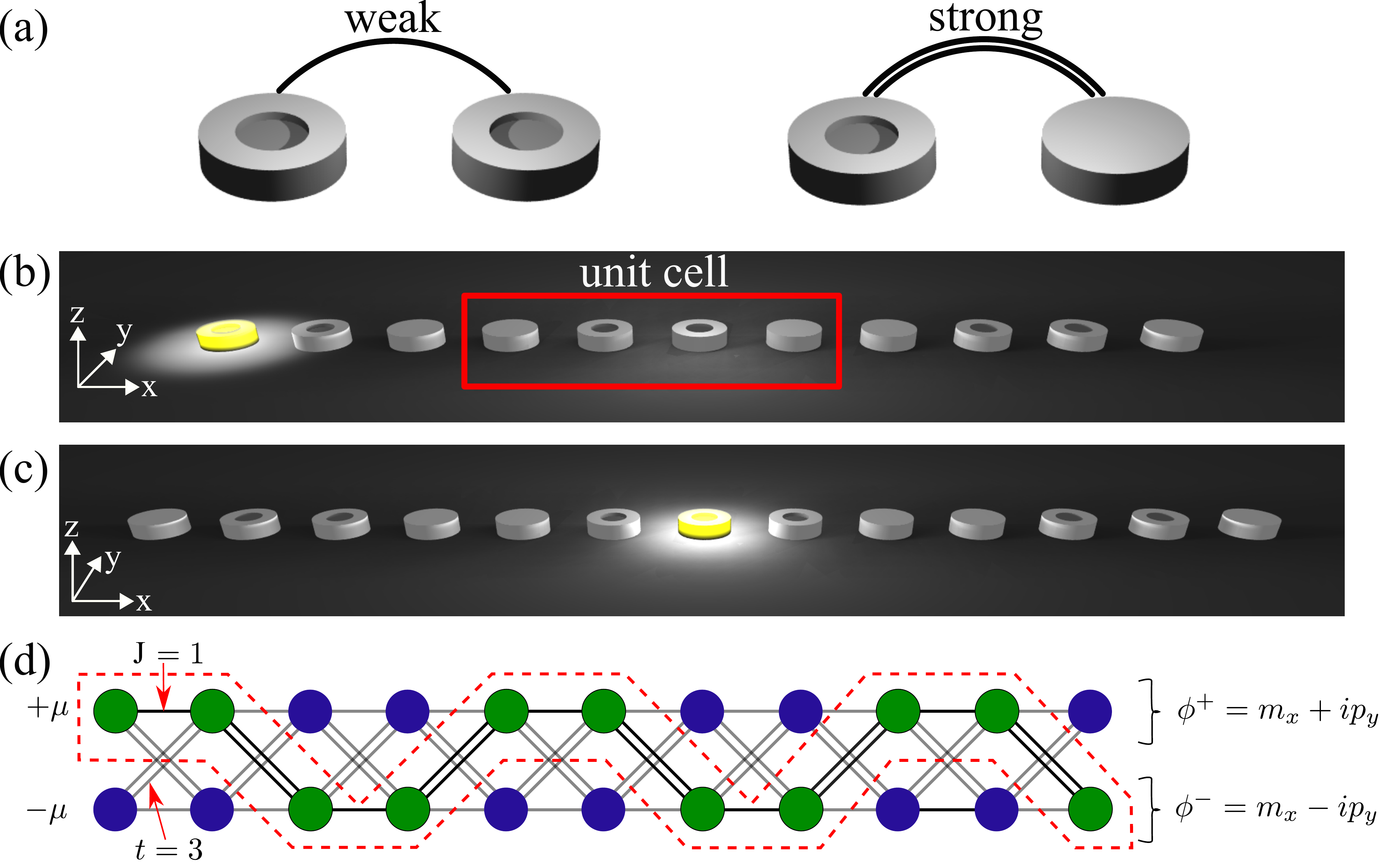}
\caption{One-dimensional arrays of high-permittivity all-dielectric disks with staggered bianisotropy. (a) The effective coupling between two particles with the same bianisotropy is weaker than that between the disks with the opposite signs of bianisotropy parameter. (b) Finite array of bianisotropic disks. The red frame highlights the four-element unit cell.  Topological state arises at the left edge of the array terminated at the weak bond. (c) Interface state localized at the ``weak-weak'' defect embedded in the array. (d) Mapping of the original electromagnetic problem onto the tight-binding model. Green and blue sites with $+\mu$ and $-\mu$ eigenfrequencies correspond to the modes comprised of orthogonal electric and magnetic dipoles with $\pm\pi/2$ phase difference.}
\label{fig:Structure}
\end{center}
\end{figure}

In this work we take a conceptually different approach and uncover an avenue to engineer photonic topological states via the {\it staggered bianisotropy pattern.} 
Our proposal is based on the general observation that the effective coupling between the meta-atoms with the same bianisotropy is typically weaker than that between the elements with the opposite signs of bianisotropy. 
While the suggested approach is quite univeral and can be potentially applied in a variety of platforms, including metamaterials, polariton micropillars, waveguide arrays and all-dielectric structures, here we focus on the specific implementation based on high-index dielectric disks with bianisotropic response ensured by mirror symmetry breaking~\cite{Alaee-Rockstuhl}. 
In this case, the sign of bianisotropy and the magnitude of the coupling parameter can easily be altered by flipping the disk, as illustrated in Fig.~\ref{fig:Structure}(a). 
Exploiting such strategy, we design the arrays depicted in Figs.~\ref{fig:Structure}(b,c) featuring an alternating pattern of effective coupling constants and supporting the topological modes at the edge and interface. 
% Furthermore, the interface of two such arrays with different dimerizations [Fig.~\ref{fig:Structure}(c)] hosts the interface states with frequency close to that of the edge states. 
To prove the topological origin of these states, we employ generic coupled dipole model and map the electromagnetic problem onto the tight-binding system sketched in Fig.~\ref{fig:Structure}(d) which in the limiting case can be reduced to the well-celebrated Su-Schrieffer-Heeger model~\cite{Su}.

\section{Results}
\subsection{Theoretical model}\label{sec:Theory}

The physics behind the topological states in our setup can be qualitatively understood based on the discrete dipole approach, when the dipole moments ${\bf d}_n$ and ${\bf m}_n$ of the particles comprising the array are related to the external electromagnetic fields ${\bf E}_n$ and ${\bf H}_n$ as 
\begin{gather}
\textbf{d}_n = \hat{\alpha}^{\rm{ee}} \textbf{E}_n + \hat{\alpha}_n^{\rm{em}} \textbf{H}_n\:,\label{dmoment}\\	
\textbf{m}_n = \hat{\alpha}_n^{\rm{me}} \textbf{E}_n + \hat{\alpha}^{\rm{mm}} \textbf{H}_n\:,\label{mmoment}
\end{gather}
where $\hat{\alpha}_n^{\rm{em}}$ and $\alpha_n^{\rm{me}}$ are $3\times 3$ polarizability tensors capturing the   bianisotropic response of $n^{\rm{th}}$ particle. Spatial symmetries of the disk together with the time-reversal symmetry restrict the structure of bianisotropy tensors as follows:
\begin{equation}\label{Beta}
\hat{\alpha}_n^{\rm{em}} = \hat{\alpha}_n^{\rm{me}} = 
\begin{pmatrix}
0 & i\beta_n & 0\\
-i\beta_n & 0 & 0\\
0 & 0 & 0
\end{pmatrix}
\:,
\end{equation}
where all $\beta_n$ have the same absolute value, but the sign differs depending on $n$. Electric and magnetic polarizability tensors have a diagonal form, 
\begin{equation}\label{Alpha}
\hat{\alpha}^{\rm{ee},\rm{mm}}
= \begin{pmatrix}
\alpha^{\rm{ee,mm}}_{\bot} & 0 & 0\\
0 & \alpha^{\rm{ee,mm}}_{\bot} & 0\\
0 & 0 & \alpha_{zz}^{\rm{ee},\rm{mm}}
\end{pmatrix}.
\end{equation}
Here, we assume that $z$ axis is the axis of the disk. We also focus on the frequency range near the resonance of in-plane dipole moments. Therefore, $z$-oriented dipoles are off-resonant and can be neglected for simplicity.

The near fields ${\bf E}_n$ and ${\bf H}_n$ causing the polarization of a given disk in the nearest-neighbor approximation are defined as
\begin{gather}
{\bf E}_n=\sum\limits_{l=n\pm 1}\,\left\lbrace\hat{G}^{\rm{ee}}({\bf{r}}_n-{\bf{r}}_l)\,\textbf{d}_l + \hat{G}^{\rm{em}}(\textbf{r}_n-{\bf r}_l)\,\textbf{m}_l\right\rbrace\:,\\
{\bf H}_n =\sum\limits_{l=n\pm 1}\,\left\lbrace \hat{G}^{\rm{me}}(\textbf{r}_n-{\bf r}_l)\,\textbf{d}_l + \hat{G}^{\rm{mm}}(\textbf{r}_n-{\bf r}_l)\,\textbf{m}_l\right\rbrace\:,
\end{gather}
where $\hat{G}({\bf r})$ denote the dyadic Green's functions~\cite{Novotny}. To simplify the analysis further, we keep only the near field terms in the expressions for dyadic Green's functions. This yields $\hat{G}^{\rm{em}}=-\hat{G}^{\rm{me}}\approx 0$, $G^{\rm{ee}}_{xx}=G^{\rm{mm}}_{xx}=2/a^3$, $G^{\rm{ee}}_{yy}=G^{\rm{mm}}_{yy}=-1/a^3$, where $a$ is the distance between the adjacent particles and $x$ axis is aligned with the axis of the array. Combining Eqs.~\eqref{Beta}, \eqref{Alpha}, we construct an inverse polarizability tensor
\begin{equation}
\hat{\alpha}^{-1}=\begin{pmatrix}
\hat{\alpha}^{\rm{ee}} & \hat{\alpha}^{\rm{em}}\\
\hat{\alpha}^{\rm{me}} & \hat{\alpha}^{\rm{mm}}
\end{pmatrix}^{-1}
=
\begin{pmatrix}
u & 0 & 0 & -i\,v\\
0 & u & i\,v & 0\\
0 & -i\,v & u & 0\\
i\,v & 0 & 0 & u
\end{pmatrix}\:.
\end{equation}
Here, $u=\alpha_{\bot}/(\alpha_{\bot}^2-\beta^2)$ and $v=\beta/(\alpha_{\bot}^2-\beta^2)$. To obtain a tractable eigenvalue problem, we assume that $v$ parameter is independent of $\om$ in the frequency range of interest, while $u=(\om-\om_0)/A$ exhibits a resonant behavior such that electric and magnetic dipole resonances of the disk  overlap at frequency $\omega_0$ for in-plane dipoles. Finally, we switch to the basis constructed from the eigenmodes of an isolated disk
\begin{equation}
\ket{\psi}=\left(p_x+i\,m_y, p_x-i\,m_y, m_x+i\,p_y, m_x-i\,p_y\right)^T\:.
\end{equation}
As a result, we derive the following eigenvalue equation
\begin{equation}\label{EigEquation}
\begin{pmatrix}
\eps-\mu_n & 0\\
0 & \eps+\mu_n
\end{pmatrix}
\,
\begin{pmatrix}
\phi_n^{(+)}\\
\phi_n^{(-)}
\end{pmatrix}
=
\begin{pmatrix}
1 & 3\\
3 & 1
\end{pmatrix}\,
\begin{pmatrix}
\phi_{n-1}^{(+)}+\phi_{n+1}^{(+)}\\
\phi_{n-1}^{(-)}+\phi_{n+1}^{(-)}
\end{pmatrix}\:,
\end{equation}
where $\eps=2\,(\om-\om_0)/A$ plays the role of energy variable, $\mu_n=2\,v_n\,a^3$ is a dimensionless parameter which quantifies particle bianisotropy, $\phi_n^{(\pm)}$ is either $p_{nx}\pm i\,m_{ny}$ or $m_{nx}\pm i\,p_{ny}$. 

Note, due to the assumed dual symmetry of the meta-atoms $(p_x, m_y)$ modes are described by the same effective Hamiltonian as $(p_y, m_x)$ ones. That is why we focus on a single $(p_y, m_x)$ polarization below.

Eigenvalue problem Eq.~\eqref{EigEquation} corresponds to the tight-binding model depicted in Fig.~\ref{fig:Structure}(d). Importantly, the coupling between $\phi_n^{(+)}$ and $\phi_{n+1}^{(-)}$ is three times larger than the coupling between $\phi_n^{(+)}$ and $\phi_{n+1}^{(+)}$. In the limit $\mu\gg 1$, i.e. when the disks are almost non-interacting, all modes of the system have either $\eps=+\mu$ or $\eps=-\mu$. Interaction splits these modes into the bands centered around $+\mu$ and $-\mu$. The modes with $\eps\approx\mu$ are mostly localized in the sites of Fig.~\ref{fig:Structure}(d) highlighted by green with eigenfrequency $+\mu$. Ignoring for the moment the rest of the sites, we immediately obtain a series of alternating coupling links as in paradigmatic Su-Schrieffer-Heeger model~\cite{Su}, which is known to support the topological state at the weak link edge.

Based on this simple reasoning, we predict the topological state at the edge of our array where the first two  particles have the same sign of magneto-electric coupling [the left edge in Fig.~\ref{fig:Structure}(b)]. Below, we confirm this  prediction both by solving the analytical model Eq.~\eqref{EigEquation} and by performing full-wave numerical simulations for the realistic design.

\subsection{Topological system design}

\begin{figure}[b]
\begin{center}
\includegraphics[width=1.0\linewidth]{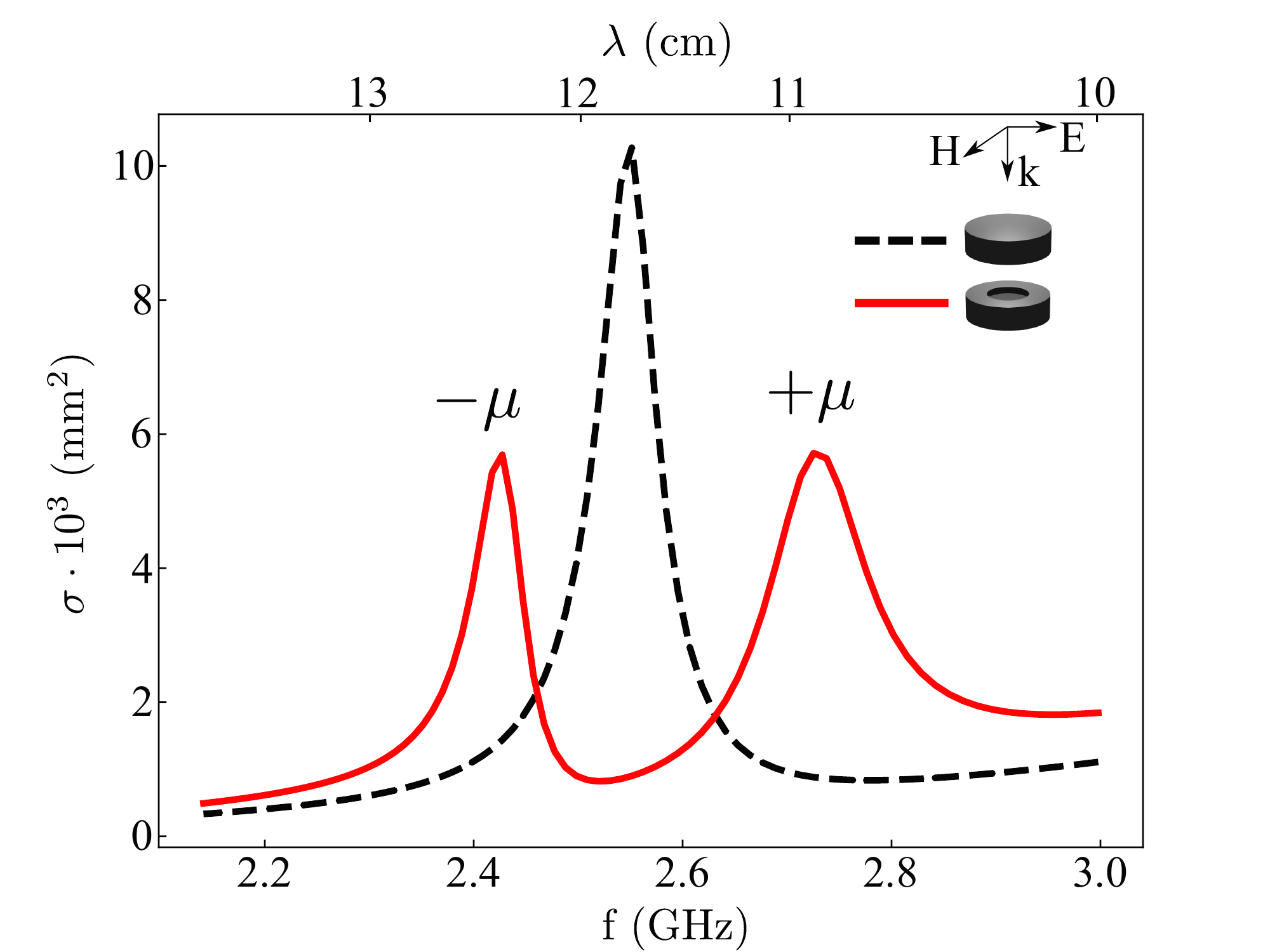}
\caption{Scattering spectra of the disks illuminated by the plane wave from the top as illustrated in inset. Dashed black line corresponds to the non-perturbed disk with diameter $D_0=27.5$~mm and height $H_0=11.0$~mm, whereas solid red curve corresponds to the perturbed disk with diameter $D = 29.1$~mm and height $H = 11.607$~mm, containing the notch with diameter $d = D/2=14.55$~mm and depth $h = H/4$. Both disks are made of the high-index material with permittivity $\eps=39$.}
\label{fig:Disk}
\end{center}
\end{figure}

As a specific platform to implement the proposed system we use dielectric disks made from commercially available high-permittivity ceramics with $\eps=39$ in $1-3$~GHz frequency range. By tuning the aspect ratio of the disk, we first ensure that electric and magnetic dipole resonances for in-plane dipoles overlap.

We test the response of the designed disk to the incident plane wave in the frequency range $2.2-3.0$~GHz (Fig.~\ref{fig:Disk}). Electric and magnetic dipoles provide the dominant contributions to the multipolar expansion of the scattered fields~\cite{Corbaton}, as shown in Fig.~S2 of Supplementary Materials. Hence, dipole model should adequately describe the response of the disk in the studied frequency range which is bounded from the bottom by the frequency of magnetic dipole resonance for $z$-oriented dipoles, $f_{\rm{b}}=1.776$~GHz, and from the top by electric dipole resonance for $z$-oriented dipoles, $f_{\rm{t}}=2.810$~GHz.

\begin{figure}[b]
\begin{center}
\includegraphics[width=0.97\linewidth]{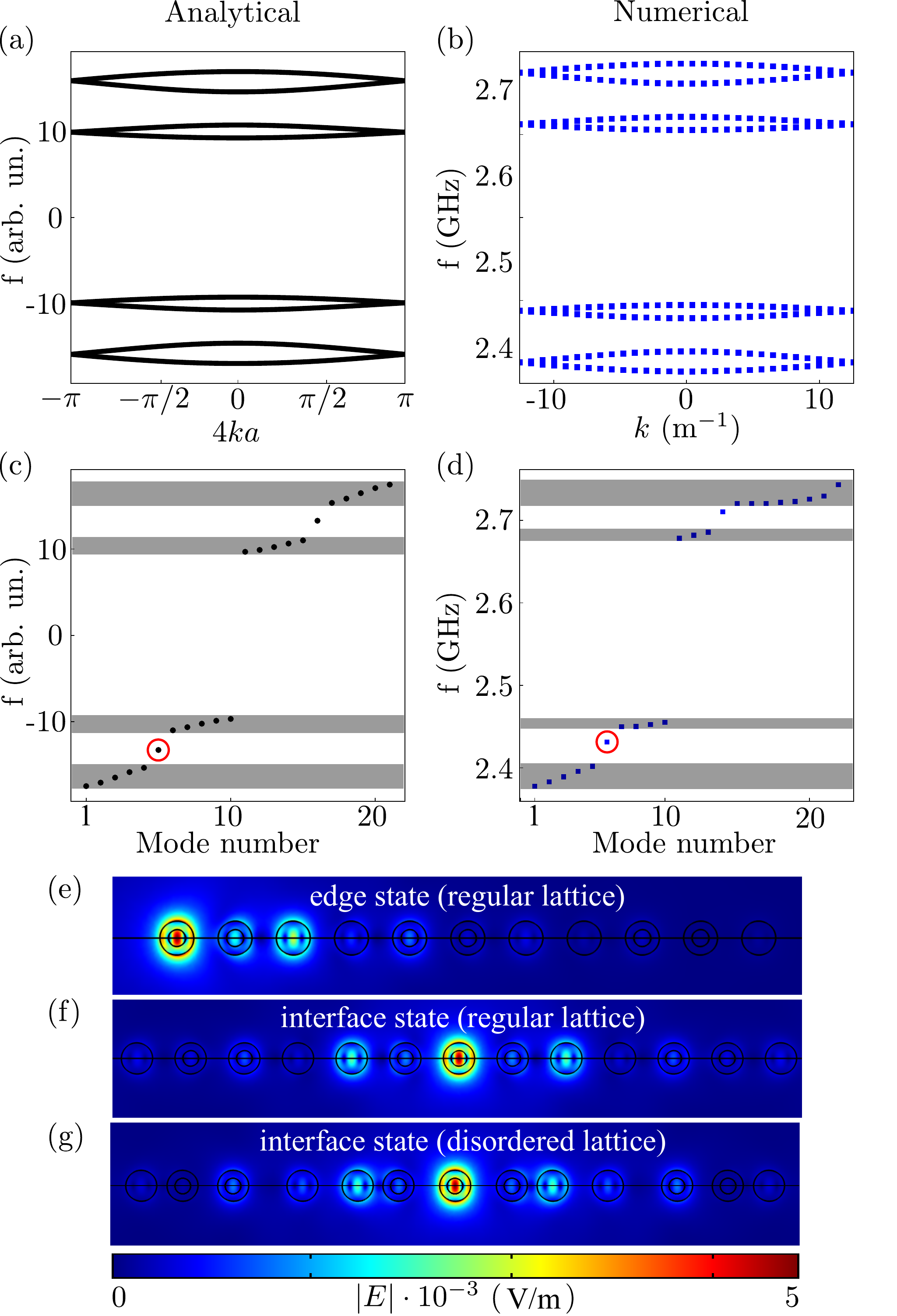}
\caption{Bandstructure and eigenmodes. (a) Tight-binding calculation of a bulk spectrum ($\mu = 13$) in a periodic array. (b) Full-wave numerical simulations of the dispersion of dipole modes in the array of dielectric disks with a period $4\,a=20$~cm performed in Comsol Multiphysics software package. Only modes with $y$-oriented electric dipoles are shown. Parameters of the disk are the same as in Fig.~\ref{fig:Disk}. (c,d) Eigenmodes of the finite array consisting of $N = 11$ disks obtained via (c) analytical model, Eq.~\eqref{EigEquation}; (d) full-wave numerical simulations with the distance between the adjacent disks $a=5$~cm. Shaded domains indicate the boundaries of the bulk bands. Eigenmode with circled frequency $f=(2.4373+0.0356\,i)$~GHz corresponds to the topological edge state with near field profile $|{\bf E}|$ in $z=0$ plane shown in panel (e). (f) Field profile $|{\bf E}|$ in $z=0$ plane of topological interface state with frequency $f=(2.4417+0.0356\,i)$~GHz localized at the weak-weak defect. (g) Simulated field profile of the interface state with frequency $f=(2.4395+0.0356\,i)$~GHz supported by the disordered array with the standard deviation in the disk position $\sigma=1.0$~cm, i.e. $\sigma/a=0.2$.}
\label{fig:Modes}
\end{center}
\end{figure}
In the case of non-perturbed disk, the scattering spectrum features a single characteristic peak corresponding to the overlapping electric and magnetic dipole resonances. Bianisotropy of the disk causes the hybridization of electric and magnetic dipole modes and results in two peaks in the scattering spectrum with the splitting between them proportional to bianisotropy parameter: $\Delta\eps=2\,\mu$. In our case, the frequency splitting reaches $300$~MHz thus ensuring the sufficient magnitude of bandgaps existing in the array of such disks. 

With the chosen design of the individual meta-atom, we now investigate the states supported by the periodic array. Since the unit cell of our system contains four disks and each disk has two degrees of freedom for a given polarization, $p_y$ and $m_x$, a periodic array has eight dipole bands shown in Fig.~\ref{fig:Modes}(a). Note that each pair of the bands touches at the boundary of Brillouin zone $k=\pm\pi/(4\,a)$. This degeneracy is associated with the additional symmetry of the model, namely, the invariance under reflection of the array in $Oxy$ plane followed by the unit cell shift by half a period. These results of analytical model appear to be in a good agreement with the results of full-wave numerical simulations shown in Fig.~\ref{fig:Modes}(b).

Remarkably, the spectrum of the system in both cases [Fig.~\ref{fig:Modes}(a,b)] is symmetric with respect to central frequency which hints to a chiral symmetry of our system. As we explicitly show in Supplementary Materials, our model Eq.~\eqref{EigEquation} indeed possesses chiral symmetry and therefore Bloch Hamiltonian can be brought to the form
\begin{equation}
\hat{H}(k)=
\begin{pmatrix}
0 & \hat{Q}(k)\\
\hat{Q}^\dag(k) & 0
\end{pmatrix}
\end{equation}
with $\text{det}\,\hat{Q}(k)=\mu^4-128\,\cos k+128$. Thus, the winding number appears to be zero and zero-energy topological states are absent.

Nevertheless, the system still supports the topological edge states. To demonstrate this, we calculate the spectrum of a finite array of $N=11$ disks shown in Fig.~\ref{fig:Structure}(b) both with our analytical model [Fig.~\ref{fig:Modes}(c)] and using full-wave numerical simulations [Fig.~\ref{fig:Modes}(d)].

Our analytical model predicts that the edge states localized at the left edge of the array arise at two frequencies: in the gap between the second and the third band and also between the sixth and seventh bands counting from the bottom. Due to chiral symmetry of the system, these states have symmetric frequencies with respect to ``zero-energy'' level. To provide a clear demonstration of the topological origin of these edge states, we further elaborate our qualitative reasoning from Sec.~\ref{sec:Theory} deriving an effective Hamiltonian for the upper and lower groups of four bands taking the limit of large $\mu$ and using degenerate perturbation theory~\cite{Bir} with respect to $1/\mu$ parameter. As we prove, each of these groups of bands is described by the effective Su-Schrieffer-Heeger Hamiltonian with the effective next-nearest-neighbor coupling, and the energy of the edge states is given by the approximate formula $\eps_{\rm{edge}}^{(\pm)}=\pm\mu\pm 4/\mu$. Hence, our analytical model yields a pair of topological edge states for each of polarizations.

However, in the case of a realistic design based on bianisotropic disks, only lower-frequency topological state persists, while higher-frequency edge state is quite poorly localized being close to the edge of the bulk band [Fig.~\ref{fig:Modes}(d)]. The main reason for that is the influence of electric dipole resonance for $z$ oriented dipoles, which is spectrally quite close making inapplicable the developed theoretical model.

Thus, we focus on the lower-frequency topological edge state. Its near field profile shown in Fig.~\ref{fig:Modes}(e) corresponds to that expected for SSH model. Examining the interface of the two SSH arrays in geometry Fig.~\ref{fig:Structure}(c), we recover a topological interface state existing at frequency very close to that of the edge state with the near field profile shown in Fig.~\ref{fig:Modes}(f).

\begin{figure}[t]
\begin{center}
\includegraphics[width=1.0\linewidth]{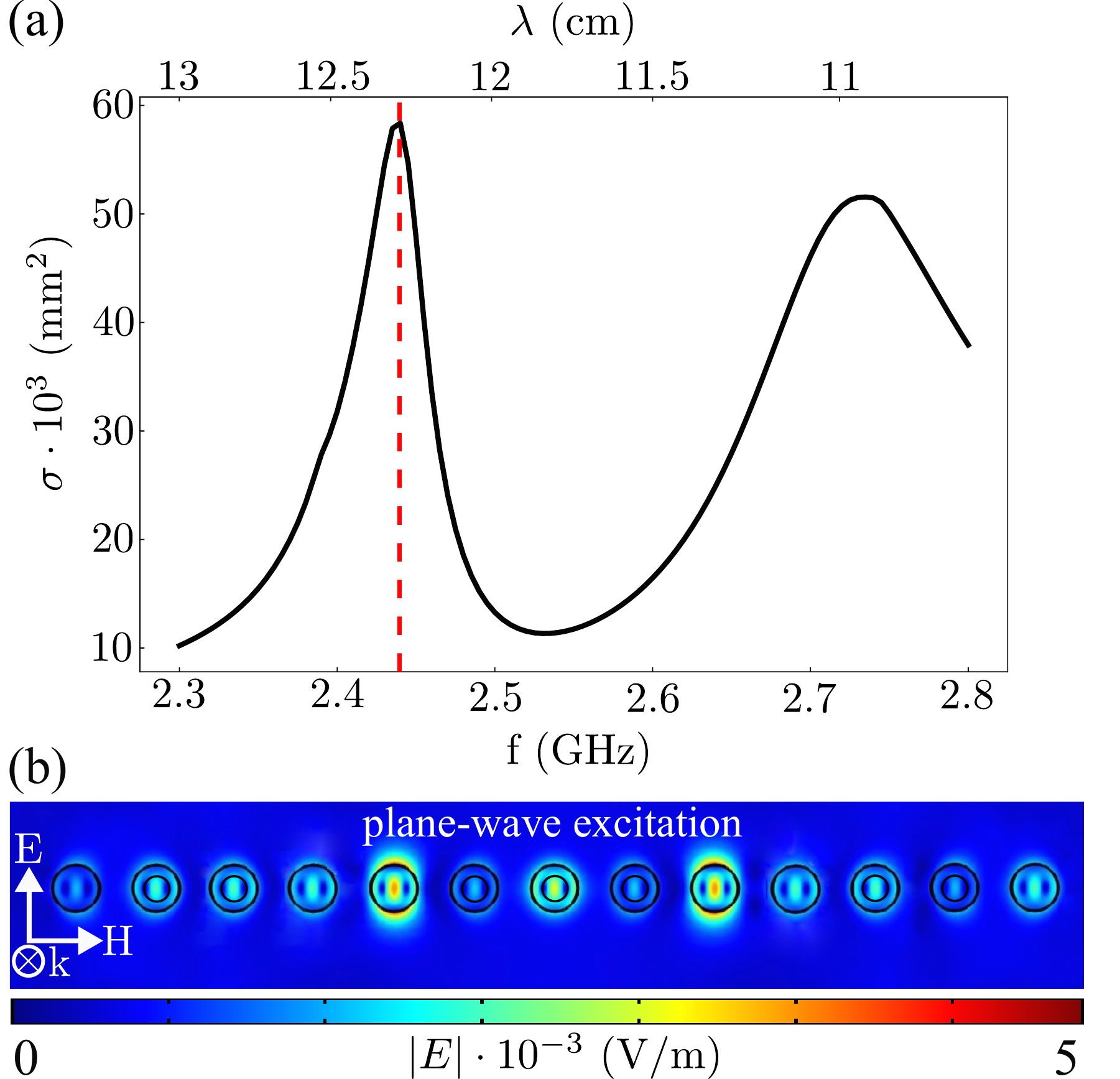}
\caption{(a) Scattering spectrum of the array composed of bianisotropic disks in geometry of Fig.~\ref{fig:Structure}(c). (b) Near field distribution $|{\bf E}|$ in $z=0$ plane at frequency $f=2.445$~GHz of lower-frequency peak marked by the vertical dashed line in panel (a).}
\label{fig:ScatteringArray}
\end{center}
\end{figure}

To illustrate the topological protection of the interface state, we simulate the disordered geometry of the array with the standard deviation in the distance between the disks $\sigma$ reaching 20\% of the average distance between the particles. In this case, the interface mode frequency remains almost unchanged, and the field distribution depicted in Fig.~\ref{fig:Modes}(g) strongly resembles that in the periodic case, Fig.~\ref{fig:Modes}(f).

To further probe the topological interface state, we consider the scattering of the incident plane wave on the array with the weak-weak defect in the middle in geometry of Fig.~\ref{fig:Structure}(c). Similarly to the case of a single disk, the scattering spectrum depicted in Fig.~\ref{fig:ScatteringArray}(a) features two characteristic peaks. Interestingly, the lower-frequency peak appears to be quite narrow, and the corresponding field distribution [Fig.~\ref{fig:ScatteringArray}(b)] resembles that of the topological interface mode. This demonstrates the possibility to excite topological interface states selectively.

\section{Conclusions}

To sum up, in this Article we have put forward a concept of topological states mediated by staggered  bianisotropy. While the traditional time-reversal-invariant designs employ spatially uniform profile of magneto-electric coupling, we demonstrate that spatially varying bianisotropic response provides an alternative route to engineer the topological states providing an easy access to dynamically reconfigure them by inverting the sign of bianisotropy. We believe that our approach to tailor the topological states is very general and can be extended towards two-dimensional topological insulators as well as higher-order topological structures. Finally, applying the developed concept to waveguide arrays, we expect that our system can be implemented not only at microwave frequencies but also in the visible range.

\section*{Acknowledgements}
We acknowledge valuable discussions with Kseniia Baryshnikova. Theoretical models were supported by the Russian Science Foundation (Grant No.~16-19-10538), numerical simulations were supported by the Russian Foundation for Basic Research (Grant No.~18-32-20065). M.A.G. acknowledges  partial support by the Foundation for the Advancement of Theoretical Physics and Mathematics ``Basis''. D.A.S. acknowledges funding from the Australian Research Council
Early Career Researcher Award (DE190100430) and the Russian Foundation for Basic Research (Grants No.~18-02-00381, 19-02-00261).

%\section*{Conflict of interest}
%The authors declare no conflict of interest.
%
%\section*{Keywords}
%Topological photonics, edge states, bianisotropy, Mie resonances.

% The authors' biographies are given inside an \authorbox command 
% each, all together enclosed in this environment.  Here is the 
% syntax:  
% \authorbox{<name>}{Authorname}{Text}
% where "<name>" is the Filename of the portrait picture, if any; it
% should consist of the author's last name prefixed with "cv".
% "Authorname" is this respective author's name being bolded as the
% first word(s) of the biography text, the rest of which follows in
% the third argument.
%\begin{biographies}
%  \authorbox{cv}{}{}
%  \authorbox{cv}{}{}
%  \authorbox{cv}{}{}
%\end{biographies}
% 
% Give your bibliography below.  If you are using BibTeX please make 
% sure to enclose all .bib files needed!

%\bibliographystyle{lpr}
\bibliography{TopologicalLib}
%\begin{thebibliography}{00}
%\bibitem{1} A. Uthor, A. N. Otherauthor, and Th. Irdauthor,
%  Laser Photonics Rev. \textbf{1}, 33--45 (2007).
%\end{thebibliography}

\end{document}

\newcommand{\e}{{\rm e}}
\newcommand{\norm}[1]{\left\lVert#1\right\rVert}
\newcommand{\rmi}{{\rm i}}
\renewcommand{\Im}{\mathop\mathrm{Im}\nolimits}
\newcommand{\red}[1]{{\color{red}#1}}
\newcommand{\blue}[1]{{\color{blue}#1}}
\newcommand{\comment}[1]{{\color{red}{\it ~Maxim:~}\tt #1}}

\renewcommand{\cite}[1]{[\onlinecite{#1}]}

\newcommand{\eps}{\varepsilon}      %Greek epsilon
\newcommand{\om}{\omega}      %Greek omega
\newcommand{\kap}{\varkappa}      %Greek kappa
\newcommand{\skvv}[2]{\left<#1\left|#2\right.\right>} %scalar products

\renewcommand{\thefigure}{S\arabic{figure}}
\renewcommand{\theequation}{S\arabic{equation}}
\title{Photonic topological states mediated by staggered bianisotropy.\\Supplementary Materials}
\author{Daniel~A.~Bobylev}
\affiliation{ITMO University, Saint Petersburg 197101, Russia}
\author{Daria~A.~Smirnova}
\affiliation{Nonlinear Physics Centre, Australian National University, Canberra ACT 2601, Australia}
\affiliation{Institute of Applied Physics, Russian Academy of Science, Nizhny Novgorod 603950, Russia}
\author{Maxim~A.~Gorlach}
\affiliation{ITMO University, Saint Petersburg 197101, Russia}
\email{m.gorlach@metalab.ifmo.ru}
\maketitle
\tableofcontents

\ \

In these Supplementary Materials, we provide further details on our theoretical model and numerical simulations of a single disk scattering spectrum.

\section{Bloch Hamiltonian}\label{sec:Bloch}

We start our analysis here from Eq.~(9) of the article main text and choose the unit cell to be inversion-symmetric as depicted in Fig.~1(b) in the main text. Additionally, we consider a fixed polarization identifying $\phi^{(\pm)}$ with $m_x\pm i\,p_y$.

The periodic part of the wave function is defined as $\ket{u_k}=\left(\phi_1^{(+)},\phi_1^{(-)},\phi_2^{(+)},\phi_2^{(-)},\phi_3^{(+)},\phi_3^{(-)},\phi_4^{(+)},\phi_4^{(-)}\right)^T$. The resulting $8\times 8$ Bloch Hamiltonian reads:
\begin{equation}\label{BlochHamiltonian}
\hat{H}(k)=
\begin{pmatrix}
\mu & 0 & 1 & 3 & 0 & 0 & e^{-ik} & 3\,e^{-ik}\\
0 & -\mu & 3 & 1 & 0 & 0 & 3\,e^{-ik} & e^{-ik}\\
1 & 3 & -\mu & 0 & 1 & 3 & 0 & 0\\
3 & 1 & 0 & \mu &3 & 1 & 0 & 0\\
0 & 0 & 1 & 3 & -\mu & 0 & 1 & 3\\
0 & 0 & 3 & 1 & 0 & \mu & 3 & 1\\
e^{ik} & 3\,e^{ik} & 0 & 0 & 1 & 3 & \mu & 0\\
3\,e^{ik} & e^{ik} & 0 & 0 & 3 & 1 & 0 & -\mu
\end{pmatrix}\:.
\end{equation}
Thus, the designed array has 8 bands corresponding to the dipole excitations with a given polarization $(p_y, m_x)$.

\section{Pairwise degeneracy of the bands at the edge of Brillouin zone}\label{sec:Degeneracy}

In our calculations we observe that each pair of the bulk bands becomes degenerate at the edge of Brillouin zone for $k=\pm K\equiv\pm\pi/(4\,a)$ [see Fig.~3(a) of the article main text]. As we show in this section, this property is related to the symmetry of the system under mirror reflection in $Oxy$ plane accompanied by the unit cell shift by half a period.

Reflection in $Oxy$ plane preserves the magnitude of electric dipole moments which are polar vectors, and changes the sign of magnetic dipole moments which are axial vectors: $\tilde{m}_{nx}=-m_{nx}$, $\tilde{p}_{ny}=p_{ny}$. Therefore, the components $\phi_n^{(\pm)}$ transform as $\tilde{\phi}_n^{(+)}=-\phi_n^{(-)}$ and $\tilde{\phi}_n^{(-)}=-\phi_n^{(+)}$. Hence, for $k=\pm K$ this symmetry operation transforms the wave function as $\ket{\tilde{\psi}}=S\,\ket{\psi}$ with the transformation matrix
\begin{equation}\label{DegeneracyMatrix}
S=\begin{pmatrix}
0 & 0 & -\sigma_x & 0\\
0 & 0 & 0 & -\sigma_x\\
\sigma_x & 0 & 0 & 0\\
0 & \sigma_x & 0 & 0
\end{pmatrix}\:,
\end{equation}
where $\sigma_x$ is $2\times 2$ Pauli matrix. It is straightforward to check that $S\,\hat{H}(K)-\hat{H}(K)\,S=0$, i.e. matrix Eq.~\eqref{DegeneracyMatrix} commutes with the Hamiltonian for $k=K$. Hence, any eigenstate of the Bloch Hamiltonian $\ket{\psi_1(K)}$ is degenerate with another eigenstate $S\,\ket{\psi_1(K)}$ corresponding to the same Bloch wave number. At the same time, $S^2=-I$, i.e. double multiplication by $S$ yields the initial eigenstate. Therefore, the symmetry described by the matrix $S$ explains pairwise degeneracy of the Bloch bands at the edge of Brillouin zone.

It should be stressed, that the degeneracy of the Bloch bands at the edge of Brillouin zone holds not only in our simplified theoretical model, but also in full-wave simulations, see Fig.~3(b) of the article main text.

\section{Chiral symmetry}\label{sec:ChiralSymmetry}

To prove chiral symmetry of our system, we construct the operator
\begin{equation}
P=\begin{pmatrix}
-\sigma_x & 0 & 0 & 0\\
0 & \sigma_x & 0 & 0\\
0 & 0 & -\sigma_x & 0\\
0 & 0 & 0 & \sigma_x
\end{pmatrix}\:.
\end{equation}
It is straightforward to check that this operator anticommutes with the Bloch Hamiltonian, i.e. $P\,\hat{H}(k)+\hat{H}(k)\,P=0$. Therefore, our system possesses chiral symmetry.

Next we calculate the eigenvectors of chiral symmetry operator and construct a basis out of them. Performing a unitary transformation
\begin{equation}
U=\frac{1}{\sqrt{2}}\,
\begin{pmatrix}
0 & 0 & 0 & 0 & 0 & 0 & -1 & 1\\
0 & 0 & 0 & 0 & 1 & 1 & 0 & 0\\
0 & 0 & -1 & 1 & 0 & 0 & 0 & 0\\
1 & 1 & 0 & 0 & 0 & 0 & 0 & 0\\
0 & 0 & 0 & 0 & 0 & 0 & 1 & 1\\
0 & 0 & 0 & 0 & -1 & 1 & 0 & 0\\
0 & 0 & 1 & 1 & 0 & 0 & 0 & 0  \\
-1 & 1 & 0 & 0 & 0 & 0 & 0 & 0  
\end{pmatrix}
\end{equation}
we obtain the Hamiltonian written in off-diagonal form
\begin{equation}
\hat{H}(k)=\begin{pmatrix}
0 & \hat{Q}(k)\\
Q^\dag(k) & 0
\end{pmatrix}
\:,
\end{equation}
where a single $4\times 4$ block is given by
\begin{equation}
\hat{Q}(k)=\begin{pmatrix}
-\mu & -2 & 0 & -2\,e^{ik}\\
4 & \mu & 4 & 0\\
0 & -2 & \mu & -2\\
4\,e^{-ik} & 0 & 4 & -\mu
\end{pmatrix}
\:.
\end{equation}
The determinant of this block $\text{det}\,\hat{Q}(k)=\mu^4-128\,\cos k+128$ remains real and positive for all $k$. Thus, winding number for our system is zero, which means that there are no zero-energy edge states~\cite{Ryu}. However, this does not mean that there are no topological states at all. In fact, our system provides an example of the situation when winding number is zero, but the topological states are present.

\section{Derivation of the effective Hamiltonian}\label{sec:EffectiveHamiltonian}

Bloch Hamiltonian has eight bands four of which have energy around $+\mu$, and the remaining four have energy around $-\mu$. In this section, we derive the effective Hamiltonian for the group of the four bands centered near energy $+\mu$. To apply the degenerate perturbation theory~\cite{Bir}, we take the limit of strong bianisotropy assuming $\mu\gg 1$.

\begin{figure}[ht]
\begin{center}
\includegraphics[width=0.7\linewidth]{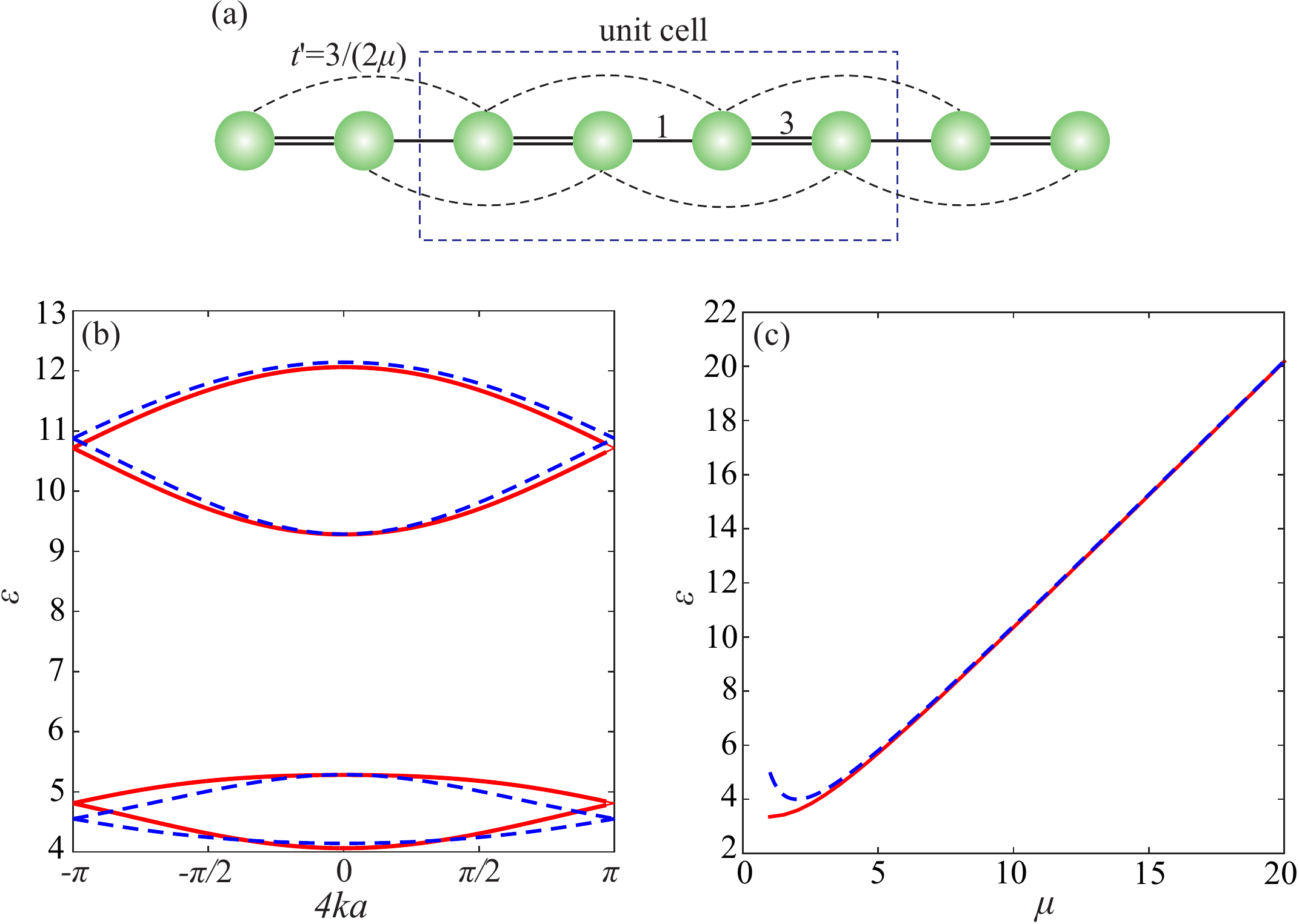}
\caption{(a) One-dimensional tight-binding model corresponding to the Bloch Hamiltonian Eq.~\eqref{EffectiveHamiltonian}. Effective next-nearest-neighbor coupling $t'=3/(2\,\mu)$ arises. (b) Calculated dispersion for the group of the four upper bands for $\mu=7$. Red solid and blue dashed lines correspond to the eigenstates of full Bloch Hamiltonian Eq.~\eqref{BlochHamiltonian} and effective Hamiltonian Eq.~\eqref{EffectiveHamiltonian}, respectively. (c) Spectral position of higher-energy edge state versus dimensionless bianisotropy parameter $\mu$. Red solid and blue dashed lines correspond to the predictions of the full model [Eq.~(9) of the article main text] and result of the first-order perturbation theory, respectively.}
\label{fig:EffHamiltonian}
\end{center}
\end{figure}

Applying the unitary transformation
\begin{equation}
U_1=\begin{pmatrix}
1 & 0 & 0 & 0 & 0 & 0 & 0 & 0\\
0 & 0 & 0 & 1 & 0 & 0 & 0 & 0\\
0 & 0 & 0 & 0 & 0 & 1 & 0 & 0\\
0 & 0 & 0 & 0 & 0 & 0 & 1 & 0\\
0 & 1 & 0 & 0 & 0 & 0 & 0 & 0\\
0 & 0 & 1 & 0 & 0 & 0 & 0 & 0\\
0 & 0 & 0 & 0 & 1 & 0 & 0 & 0\\
0 & 0 & 0 & 0 & 0 & 0 & 0 & 1\\
\end{pmatrix}\:,
\end{equation}
to the Hamiltonian Eq.~\eqref{BlochHamiltonian}, we arrive to the result
\begin{equation}
\hat{H}_1(k)=\begin{pmatrix}
\mu & 3 & 0 & e^{-ik} & 0 & 1 & 0 & 3\,e^{-ik}\\
3 & \mu & 1 & 0 & 1 & 0 & 3 & 0\\
0 & 1 & \mu & 3 & 0 & 3 & 0 & 1\\
e^{ik} & 0 & 3 & \mu & 3\,e^{ik} & 0 & 1 & 0\\
0 & 1 & 0 & 3\,e^{-ik} & -\mu & 3 & 0 & e^{-ik}\\
1 & 0 & 3 & 0 & 3 & -\mu & 1 & 0\\
0 & 3 & 0 & 1 & 0 & 1 & -\mu & 3\\
3\,e^{ik} & 0 & 1 & 0 & e^{ik} & 0 & 3 & -\mu
\end{pmatrix}
\:.
\end{equation}
We consider the terms proportional to $\mu$ as the leading-order ones, while the rest of the terms are considered as perturbation. At this point, we apply the degenerate perturbation theory~\cite{Bir}:
\begin{equation}
H_{mm'}^{\rm{eff}} = H_{mm'} + \frac{1}{2} \sum_{s} \left[ \frac{1}{E_m^{(0)}-E_s^{(0)}} + \frac{1}{E_{m'}^{(0)}-E_s^{(0)}} \right] H'_{ms} H'_{sm'},
\end{equation}
where $m,m'=1\dots 4$ and $s=5\dots 8$. As a result, we derive the following effective Hamiltonian:
\begin{equation}\label{EffectiveHamiltonian}
\hat{H}^{\rm{eff}}=\left(\mu+\frac{5}{\mu}\right)\hat{I}+\begin{pmatrix}
0 & 3 & t'\,(1+e^{-ik}) & e^{-ik}\\
3 & 0 & 1 & t'\,(1+e^{-ik})\\
t'\,(1+e^{ik}) & 1 & 0 & 3\\
e^{ik} & t'\,(1+e^{ik}) & 3 & 0\\
\end{pmatrix}\:.
\end{equation}
This Hamiltonian describes modified SSH model with the eigenfrequency of all sites equal to $\mu+5/\mu$, nearest neighbor hoppings equal to $1$ and $3$, and next-nearest-neighbor coupling $t'=3/(2\,\mu)$ depicted in Fig.~\ref{fig:EffHamiltonian}(a). Thus, finite bianisotropy induces effective next-nearest-neighbor hopping, and as a result of that chiral symmetry of the effective Hamiltonian is broken. The energies of the bands in descending order read:
\begin{gather}
\eps_1=\mu+\frac{5}{\mu}+\frac{3}{\mu}\,\cos\,\frac{k}{2}+\sqrt{10+6\,\cos\,\frac{k}{2}}\:,\\
\eps_2=\mu+\frac{5}{\mu}-\frac{3}{\mu}\,\cos\,\frac{k}{2}+\sqrt{10-6\,\cos\,\frac{k}{2}}\:,\\
\eps_3=\mu+\frac{5}{\mu}+\frac{3}{\mu}\,\cos\,\frac{k}{2}-\sqrt{10+6\,\cos\,\frac{k}{2}}\:,\\
\eps_4=\mu+\frac{5}{\mu}-\frac{3}{\mu}\,\cos\,\frac{k}{2}-\sqrt{10-6\,\cos\,\frac{k}{2}}\:.
\end{gather}
Comparing these results with the predictions of the full model Eq.~\eqref{BlochHamiltonian}, we observe quite good agreement in the region of $\mu\geq 7$ [Fig.~\ref{fig:EffHamiltonian}(b)]. Still, even at such $\mu$ breaking of chiral symmetry of the effective Hamiltonian is clearly observable.

To discuss the topological properties of the arising edge states, we present the derived Hamiltonian Eq.~\eqref{EffectiveHamiltonian} in extended band representation, keeping just two sites in the unit cell, since the intrinsic periodicity of the effective model Fig.~\ref{fig:EffHamiltonian}(a) is just $2\,a$. This yields $2\times 2$ effective Bloch Hamiltonian
\begin{equation}\label{EffHam2}
\hat{H}^{\rm{ext}}=\left(\mu+\frac{5}{\mu}+\frac{3}{\mu}\,\cos k\right)\hat{I}+
\begin{pmatrix}
0 & 1+3\,e^{-ik}\\
1+3\,e^{ik} & 0
\end{pmatrix}\:.
\end{equation}
Note that the only difference from the canonical SSH model is in the nonzero diagonal entries which provide just the $k$-dependent shift of energy bands but do not modify the structure of the eigenstates. Hence, the Zak phase is still quantized and equal to $\pi$ once the unit cell with the weak link inside is chosen.

Next we derive the asymptotics for the topological edge state arising in our model. In the case of a semi-infinite array the Hamiltonian of the effective model reads:
\begin{equation}
\hat{H}=\begin{pmatrix}
0 & 1 & t' & 0 & 0 & \dots\\
1 & 0 & 3 & t' & 0 & \dots\\
t' & 3 & 0 & 1 & t' & \dots\\
0 & t' & 1 & 0 & 3 & \dots\\
0 & 0 & t' & 3 & 0 & \dots\\
&  & \dots  & \ldots
\end{pmatrix}
\end{equation}
We treat the part proportional to $t'$ as a perturbation $\hat{V}$ and apply the first-order perturbation theory. An unperturbed wave function, describing the edge state in SSH model is given by
\begin{equation}
\ket{\psi_0}=\frac{\sqrt{8}}{3}\,\left(1,0,-\frac{1}{3},0,\frac{1}{9},0,-\frac{1}{27},0,\dots\right)^T\:,
\end{equation}
while the first-order approximation for the energy of this state is
\begin{equation}
\eps_{\rm{edge}}=\mu+\frac{5}{\mu}+\left\langle\psi_0\left|\hat{V}\right|\psi_0\right\rangle\:.
\end{equation}
Straightforward calculation yields
\begin{equation}
\eps_{\rm{edge}}=\mu+\frac{4}{\mu}\:.
\end{equation}
The derived asymptotics perfectly agrees with the result of the full model as shown in Fig.~\ref{fig:EffHamiltonian}(c) starting from $\mu\geq 2$. Hence, perturbative treatment developed here, highlights important distinctive features of our system from the SSH model from one side, proving the topological origin of the observed edge state from the other.

\section{Multipolar expansion of the scattered fields for a single disk}\label{sec:Multipole}

\begin{figure}[h]
\begin{center}
\includegraphics[width=0.45\linewidth]{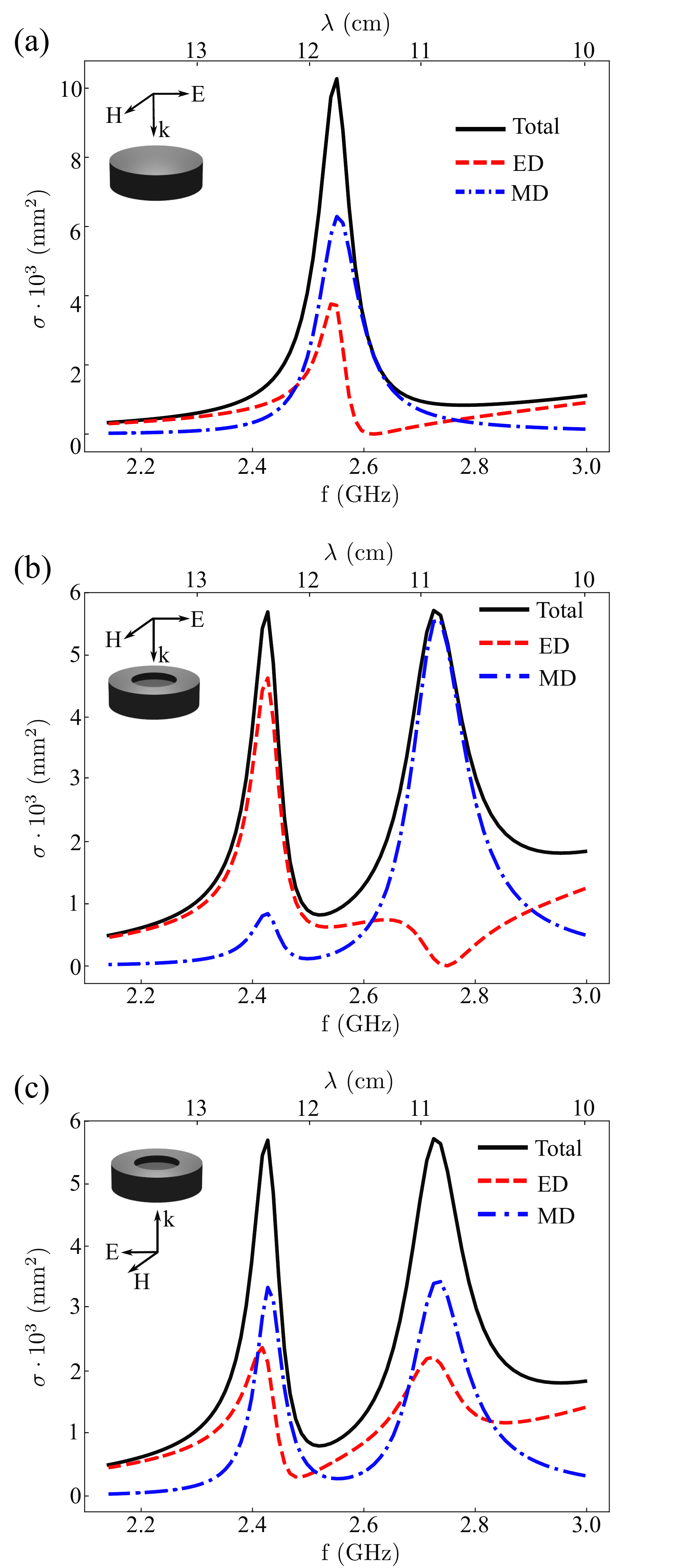}
\caption{Scattering spectrum (a) for the original non-perturbed disk with diameter and height $D_0=27.5$~mm, $H_0=11.0$~mm, respectively; (b,c) for bianisotropic ceramic disk with diameter and height $D = 29.1$~mm, $H = 11.607$~mm. Diameter and depth of the hole are $d = D/2=14.55$~mm and $h = H/4$. Top (b) and bottom (c)  illuminations are considered. Permittivity of ceramics in both cases is equal to $\eps=39$. Contributions of electric and magnetic in-plane dipole moments are shown by red dashed and blue dot-dashed lines, respectively. Frequency splitting between the peaks caused by bianisotropy is around 300~MHz.}
\label{fig:Multipole}
\end{center}
\end{figure}

To assess the validity of the discrete dipole model in our case, we simulate  scattering of a plane wave on a single disk. We start from unperturbed (i.e. symmetric) disk with the aspect ratio chosen in such a way that electric and magnetic dipole resonances for in-plane dipoles overlap [Fig.~\ref{fig:Multipole}(a)]. As expected, the scattering spectrum features a single peak. Multipolar expansion~\cite{Corbaton} indicates that electric and magnetic dipole resonances coincide, while the magnitude of electric and magnetic polarizabilities is different.

Breaking mirror symmetry of the disk, we introduce bianisotropy, which results in splitting of the central scattering peak into two side peaks. Each of them corresponds to the hybrid mode of the disk involving in general case both electric and magnetic dipole moments.

While the overall scattering cross section of the disk is the same for top and bottom illumination directions [Fig.~\ref{fig:Multipole}(b,c)], the multipolar composition of the scattered fields does depend on illumination direction~\cite{Alaee-Rockstuhl}. In both cases, however, electric dipole contribution into the higher-frequency peak has strongly asymmetric profile.

\bibliography{TopologicalLib}